\documentclass[12pt,letterpaper,preprint]{aastex}
\usepackage{graphicx}
\pdfoutput=1

\newcommand{\etal} {et al.}

\slugcomment{\it In Proceedings IAU 2009 Symposium S263, ``Icy Bodies of the
Solar System'', D.~Lazzaro, D.~Prialnik, R.~Schulz \& J.A.~Fernandez, eds.}

\shorttitle{The Dark Red Spot on KBO Haumea}

\shortauthors{Pedro Lacerda}

\begin{document}

\title{The Dark Red Spot on KBO Haumea}
\author{Pedro Lacerda}
\affil{Newton Fellow, Queen's University, Belfast BT7 1NN, United Kingdom.} 
\email{p.lacerda@qub.ac.uk} 

\begin{abstract} Kuiper belt object 136108 Haumea is one of the most
fascinating bodies in our solar system. Approximately
$2000\times1600\times1000$ km in size, it is one of the largest Kuiper belt
objects (KBOs) and an unusually elongated one for its size. The shape of Haumea
is the result of rotational deformation due to its extremely short
3.9-hour rotation period. Unlike other 1000 km-scale KBOs which are coated in
methane ice the surface of Haumea is covered in almost pure H$_2$O-ice. The
bulk density of Haumea, estimated around 2.6~g~cm$^{-3}$, suggests a more rocky
interior composition, different from the H$_2$O-ice surface.  Recently, Haumea
has become the second KBO after Pluto to show observable signs of surface
features. A region darker and redder than the average surface of Haumea has
been identified, the composition and origin of which remain unknown.  I discuss
this recent finding and what it may tell us about Haumea.
\end{abstract}


\section{Introduction}

The Kuiper belt is currently the observational frontier of our solar system.
Presumably the best kept remnants of the icy planetesimals that formed the
outer planets, Kuiper belt objects (KBOs) have been the subjects of intense
study in the past $\sim$15 years. One intriguing KBO is 136108 Haumea (formerly
2003~EL$_{61}$). First famous for its super-fast rotation and elongated shape,
Haumea went on to surprise us with a host of interesting properties.  Haumea's
spin frequency of one rotation every $\sim3.9$~hr is unparalleled for an object
this large \citep{2008ssbn.book..129S}. Its shape is rotationally deformed into
a $2000\times1600\times1000$~km triaxial ellipsoid \citep{2006ApJ...639.1238R}
to balance gravitational and centripetal accelerations. To attain such a fast
rotation, Haumea might have suffered a giant impact at the time when the Kuiper
belt was massive enough to render such events likely.  Infrared spectroscopy
has revealed a surface covered in almost pure H$_2$O ice
\citep{2007ApJ...655.1172T} which gives Haumea an optically blue colour
\citep{2007AJ....133..526T}. The surfaces of the remaining Pluto-sized KBOs
(Eris, Pluto and Makemake) are covered in CH$_4$ ice instead, granting them the
tag `Methanoids'.  Two satellites were discovered in orbit around Haumea
\citep{2006ApJ...639L..43B}, the largest of which is also coated in even purer
H$_2$O ice \citep{2006ApJ...640L..87B}. The two satellites have nearly coplanar
orbits with fast-evolving, complex dynamics due mainly to tidal effects from
the primary \citep{2009AJ....137.4766R}.  Haumea's bulk density, derived
assuming it is near hydrostatic equilibrium, is $\rho\sim2.6$~g~cm$^{-3}$
\citep{2007AJ....133.1393L}.  The surface material has density $\rho\sim1$ in
the same units implying that the interior must be differentiated and Haumea
must have more rock-rich core.  A number of KBOs showing signs of H$_2$O ice in
their surface spectra all lie close to Haumea in orbital space
\citep{2008ApJ...684L.107S}; this, plus the unusually fast spin, the
differentiated inner structure and the two small satellites also covered in
H$_2$O ice, all have been taken as evidence that Haumea is the largest remnant
of a massive collision that occured $>1$~Gyr ago
\citep{2007AJ....134.2160R,2007Natur.446..294B}.  However, several potential
members of the collisional family have been eliminated based on infrared
photometry (Snodgrass \etal, poster at this meeting).

\begin{figure}[ht]
\begin{center}
 \includegraphics[width=\textwidth]{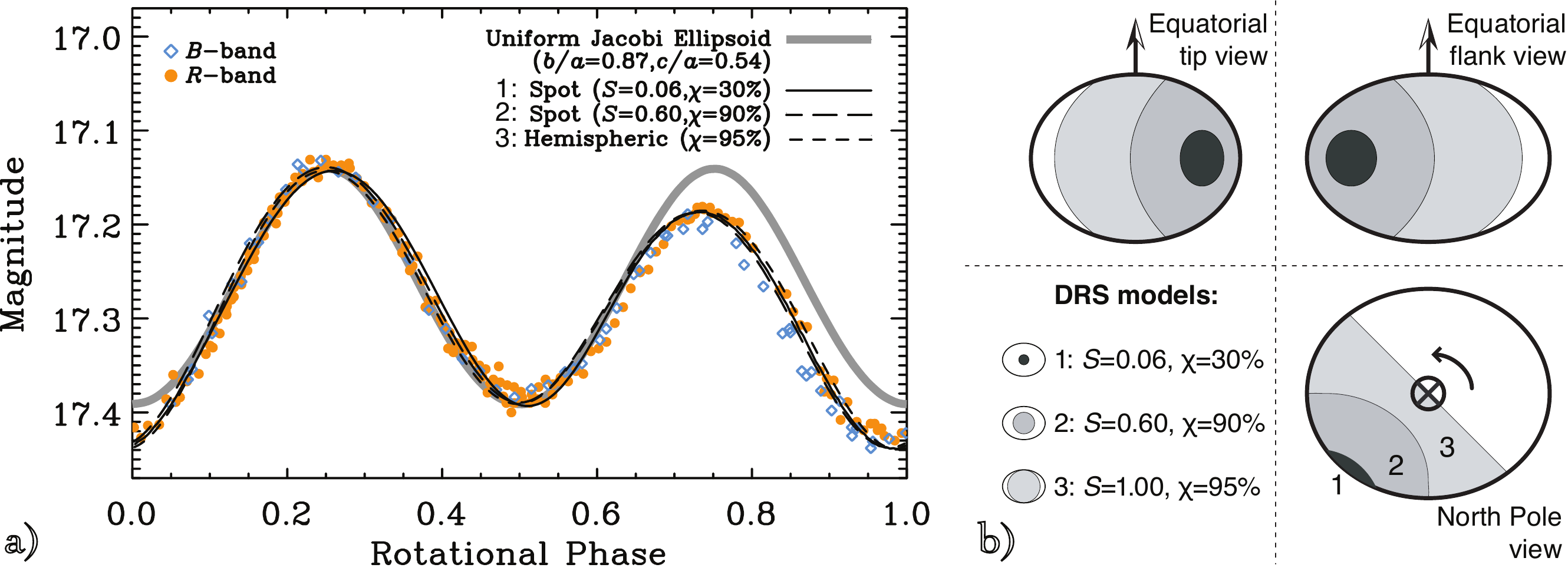} 
 \caption{\textbf{a)} Lightcurve of Haumea in two filters. Both $B$ and $R$
 data were taken over 3 nights to ensure repeatability. The effect of the dark
 red spot is apparent at rotational phases  $0.7\lesssim\phi\lesssim1.0$: the
 maximum and minimum that bracket that region appear darker and the $B$-band
 flux is consistently lower than the $R$-band flux indicating the spot is
 redder than elsewhere.  We measure a lightcurve period $P=3.9155\pm0.0001$
 hours and a photometric range $\Delta m=0.29\pm 0.02$~mag. The rotationally
 averaged colour is $B-R=0.972\pm0.017$~mag. Best fit Jacobi ellipsoid models
 are overplotted: a thick solid grey line shows how the uniform surface model
 fails to fit the dark red spot region, and thinner lines show that a small
 ($S<<1$) and very dark ($\chi<<100\%$) spot or a large ($S\sim1$) and not very
 dark ($\chi\sim100\%$) spot fit the data equally well.  \textbf{b)} Cartoon
 representation of the three spot models considered in a)
 showing the location of the spot on the surface of Haumea.  }
   \label{fig1}
\end{center}
\end{figure}

\section{The Dark Red Spot (DRS) on Haumea}

We observed Haumea in mid-2007 using the University of Hawaii 2.2m telescope
with the goal of measuring its lightcurve in two bands, $B$ and $R$
(Fig.~\ref{fig1}a). Our high-quality photometry \citep{2008AJ....135.1749L}
shows two important features: 

\begin{enumerate}

\item The lightcurve is not symmetric as would be expected from a uniform
ellipsoidal body. There is a clear asymmetry between the two sets of minima and
maxima indicating the presence of a dark region on the surface
(Fig.~\ref{fig1}a). A model lightcurve generated by placing a dark spot on the
equator of Haumea, visible at both minimum and maximum cross-section
(Fig.~\ref{fig1}b), successfully fits the data. 

\item Upon aligning the $B$ and $R$ lightcurve data we verify that the $B$
points lie consistently below the $R$ points precisely at the location of the
dark spot.  In other words, the dark spot is also redder than the average
surface.
 
\end{enumerate}

In the rest of the paper we use DRS to refer to the dark red spot.  In our
model (Fig.~\ref{fig1}) the size and relative darkness of the DRS are
degenerate: the spot may be as small as a few percent of the projected
cross-section of Haumea and be about 20\% as reflective as the rest of the
surface, or it may be as large as to take a full hemisphere of Haumea being
then only 5\% less reflective than elsewhere. The same degeneracy applies to
colour vs.\ spot size. However, assuming the DRS colour is within the range of
values typically found in the solar system, $1.0 \lesssim B-R\;(\mathrm{mag})
\lesssim 2.5$, then when directly facing the observer the spot must take
between 20\% and 60\% of the projected cross-section of Haumea, and have an
albedo between 55\% and 65\%.  This combination of colour and albedo is
consistent with, e.g.\ Eris, Makemake and the bright regions on Pluto and on
Saturn's satellite Iapetus; it is inconsistent with Pluto's darker regions,
with Pluto's satellite Charon, with Saturn's irregular satellite Phoebe and
with Centaurs Chiron and Pholus.

\begin{figure}[ht]
\begin{center} \includegraphics[width=\textwidth]{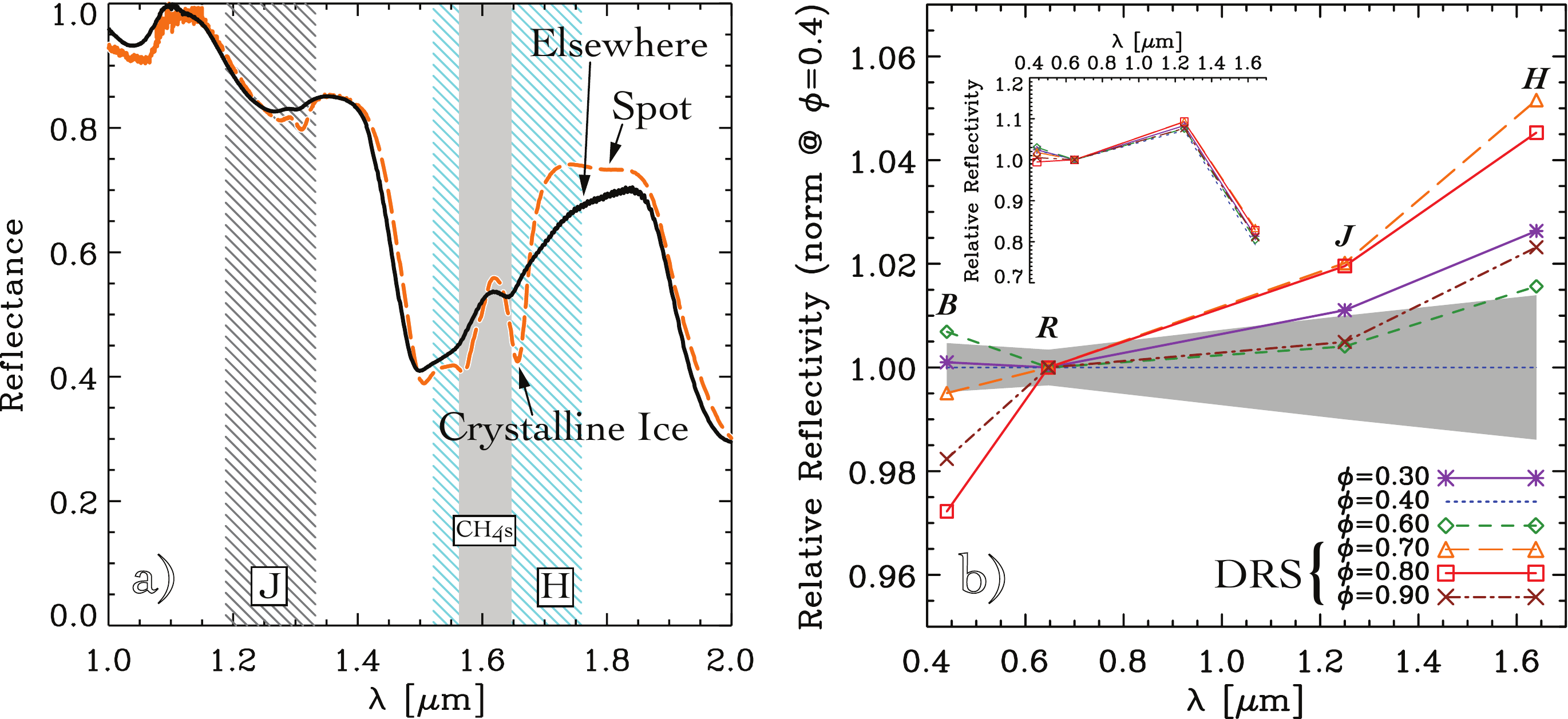} 
 \caption{\textbf{a)} Near-infrared synthetic spectra of H$_2$O-ice. The orange
 dashed line shows a spectrum of crystalline ice (indicated by the
 1.65~$\mu$m feature) while the solid black line corresponds to ice with a
 lower degree of crystallinity. \textbf{b)} Time-resolved 4-band spectrum of
 Haumea [adapted from \citet{2009AJ....137.3404L}]. Each line is a spectrum at
 a given rotational phase. At rotational phases when the DRS faces the observer
 ($0.7\lesssim\phi\lesssim1.0$) the $B$ band is depressed and the $H$ band is
 enhanced. Spectra at each rotational phase are plotted relative to $R$ band
 and all rotational phases have been normalised by $\phi=0.4$. Inset shows
 spectra before normalisation at $\phi=0.4$.  }
 \label{fig2} \end{center} \end{figure}


\section{The DRS in the infrared}

Prompted by the fact that Haumea is covered in H$_2$O ice, we set out to
investigate how the properties of the ice changed close the DRS region by
monitoring the infrared H$_2$O-ice absorption band at 1.5~$\mu$m
(Fig.~\ref{fig2}). We collected two sets of data: time-resolved,
quasi-simultaneous broadband $J$ (to probe the continuum) and medium band
CH$_4$s (to probe the 1.5~$\mu$m) data using UKIRT, and a similar dataset at
the Subaru telescope this time using broadband filters $J$ and $H$. Both
telescopes are located atop Mauna Kea. 

Neither of the filters CH$_4$s or $H$ fully probe the 1.5~$\mu$m band, and both
are affected by a narrower band at 1.65~$\mu$m which, if present, indicates
that the ice is mostly crystalline \citep{2007A&A...466.1185M}.  Our UKIRT
measurements indicate that the CH$_4$s/$J$ flux ratio decreases by a few
percent close to the DRS, while our Subaru measurement show that the $H$/$J$
flux ratio increases also by a few percent. This apparent contradiction can be
explained if the DRS is richer in crystalline H$_2$O-ice than the rest of the
surface. That would change the shape of the 1.5~$\mu$m and 1.65~$\mu$m bands
and cause exactly what is observed see Fig.~\ref{fig2}a.
\citet{2009ApJ...695L...1F} report \emph{HST} NICMOS observations of Haumea
using the broadband filters F110W and F160W centered respectively at 1.1~$\mu$m
and 1.6~$\mu$m. They find that the F160W/F110W flux ratio decreases at the DRS,
consistent with our UKIRT observations. Infrared spectra obtained at 4-m-class
telescopes show no rotational variations, likely due to lack of sensitivity
\citep{2009A&A...496..547P}.

\section{The DRS 4-band spectrum}

Using the time-resolved $BR$ data from the UH~2.2~m and $JH$ data from Subaru
we constructed 4-band spectra of Haumea as it rotates. We did this by
interpolating each of the four lightcurves and measuring the relative
differences between the bands. The result is shown in Fig.~\ref{fig2}b where we
plot 3 spectra away from the DRS ($\phi=0.3, 0.4, 0.6$) and 3 spectra close to
the DRS ($\phi=0.7, 0.8, 0.9$). The Figure shows that the DRS material is a
more efficient $B$ absorber than the rest of Haumea, hence the redder $B$$-$$R$
colour, and that it is redder in visible-to-infrared colour. In \S3 we saw that
the DRS displays bluer or redder behaviour in the $JH$ wavelength range
depending on exactly which filter bandpasses are used. 

\section{Interpretation}

It is very unlikely that the dark spot is a topographical feature such as a
mountain or valley as that would produce an achromatic change in brightness.
Instead, the spot region exhibits slight but persistent visible and infrared
colour properties that distinguish it from the rest of Haumea's surface. No
atmosphere has been detected on Haumea rendering an explanation based on
irregular condensation of gases on the surface unlikely. The fact that the DRS
absorbs $B$-band light more efficiently could indicate the presence of hydrated
minerals \citep{2007AJ....134.2046J}. Alternatively the redder tint could be
due to the presence of irradiated organic materials which would also explain
the blue behaviour observed in some infrared bandpasses
\citep{2009ApJ...695L...1F}. If confirmed, the higher degree of ice
crystallinity at the DRS could signal a recent temperature rise.

\section{Speculation}

The DRS could be a region where material from Haumea's interior is trickling
out. The high bulk density of Haumea indicates the presence of a more
mineral-rich core. If warmer, deep-lying material would find its way to the
surface it would appear darker and presumably redder than H$_2$O ice. The slight
increase in temperature and the presence of H$_2$O are useful ingredients for
mineral hydration and H$_2$O-ice crystallinization.

The DRS could also be the site of a recent impact of a $\sim1-10$~km KBO onto
Haumea. Small KBOs are dark and most are believed to be covered in red,
irradiated organic mantles. The collision would locally raise the temperature,
thereby accelerating the transition from amorphous-to-crystalline H$_2$O ice,
and the impactor material would probably leave a visible trace on the surface.

The DRS could also be due to something completely different. Time-resolved
spectroscopy of Haumea using $8-10$~m telescopes should help determine the
composition of the DRS and help solve the mystery of its origin.

\section*{Acknowledgments}

I am grateful to David Jewitt for comments on the manuscript and to the Royal
Society for the support of a Newton Fellowship.

\end{document}